**Anthropogenic influences on groundwater in the vicinity of a long-lived radioactive waste repository**

Matthew A. Thomas[1,2*], Kristopher L. Kuhlman[3], and Anderson L. Ward[4]

[1] Sandia National Laboratories, 4100 National Parks Highway, Mail Stop 1395, Carlsbad, New Mexico, 88220-9006
[2] now at U.S. Geological Survey, Box 25046, Denver Federal Center, MS 966, Denver CO 80225
[3] Sandia National Laboratories, P.O. Box 5800, Mail Stop 0747, Albuquerque, New Mexico, 87185-0747
[4] U.S. Department of Energy, 4021 National Parks Highway, Carlsbad, New Mexico, 88220-9082
[*] corresponding author; US Geological Survey, Box 25046, Denver Federal Center, MS 966, Denver, CO 80225; matthewthomas@usgs.gov; (303) 273-8588

**ABSTRACT**
The groundwater flow system in the Culebra Dolomite Member (Culebra) of the Permian Rustler Formation is a potential radionuclide release pathway from the Waste Isolation Pilot Plant (WIPP), the only deep geological repository for transuranic waste in the United States. In early conceptual models of the Culebra, groundwater levels were not expected to fluctuate markedly, except in response to long-term climatic changes, with response times on the order of hundreds to thousands of years. Recent groundwater pressures measured in monitoring wells record more than 25 m of drawdown. The fluctuations are attributed to pumping activities at a privately-owned well that may be associated with the demand of the Permian Basin hydrocarbon industry for water. The unprecedented magnitude of drawdown provides an opportunity to quantitatively assess the influence of unplanned anthropogenic forcings near the WIPP. Spatially variable realizations of Culebra saturated hydraulic conductivity and storativity were used to develop groundwater flow models to estimate a pumping rate for the private well and investigate its effect on advective transport. Simulated drawdown shows reasonable agreement with observations (average Model Efficiency coefficient = 0.7). Steepened hydraulic gradients associated with the pumping reduce estimates of conservative particle travel times across the domain by one-half and shift the intersection of the average particle track with the compliance boundary by more than two kilometers. The value of the transient simulations conducted for this study lie in their ability to (i) improve understanding of the Culebra groundwater flow system and (ii) challenge the notion of time-invariant land use in the vicinity of the WIPP.

**1.0 INTRODUCTION**

Calibration, verification, and validation are steps commonly employed to quantitatively evaluate numerical models of hydrogeologic systems. Calibration and verification rely on existing observations of the natural system. Validation refers to the ability of a model to satisfactorily predict future conditions (Zheng & Bennet, 2002). When model predictions are made on the human timescale (i.e., tens to hundreds of years), validation may be possible. However, if relevant hydrogeologic conditions extend into the geologic timescale (i.e., thousands to millions of years), direct validation through planned and controlled experiments may be impossible (Hassan, 2004). Natural analogues or natural stimuli (i.e., unplanned or uncontrolled experiments) can provide some validation information, but may not satisfy criteria for traditional model validation (Alexander et al., 2015). Even if decades of observations are available to calibrate and verify a model, the ground truth represents only a small fraction of the geologic timescale.

Long-term performance assessment modeling for radioactive waste disposal is one problem that must contend with the geologic timescale. The U.S. Department of Energy (DOE) Waste Isolation Pilot Plant (WIPP), located approximately 50 km east of Carlsbad in southeastern New Mexico, is currently the only deep geological repository for transuranic waste in the United States. The WIPP facility, shown in Figure 1, consists of surface buildings, an array of vertical shafts, and a mined repository approximately 650 m below the ground surface. The repository is situated in a thick Permian-age sequence of interbedded salt (halite) and anhydrite known as the Salado Formation. As part of compliance recertification for the WIPP, the probability of radionuclide release from the repository to the Land Withdrawal Act (LWA) boundary (i.e., the compliance boundary) inside 10,000 years is calculated with a suite of physics-based numerical models of subsurface flow and reactive solute transport (DOE, 2014).

The work presented here focuses on simulations for the Culebra Dolomite Member (Culebra) of the Permian Rustler Formation, located approximately 450 m above the repository level (see Figure 1). A conceptual schematic of the Culebra is shown in Figure 2. Groundwater flow under a natural gradient in the Culebra, mapped in Figure 3, represents a process by which radionuclides, brought from the repository elevation through inadvertent human intrusion, could reach the LWA boundary (DOE, 2014).

In early conceptual models for the Culebra used to support the Compliance Certification Application for the WIPP, groundwater levels in the vicinity of the planned repository were not expected to fluctuate, except in response to long-term changes in climatic conditions, with response times on the order of hundreds to thousands of years (Lambert and Harvey, 1987; Davies, 1989; Corbet & Knupp, 1996; Corbet, 2000). Furthermore, site licensing regulations specified that the performance of the WIPP be grounded in the extrapolation of "…current land-use practices in the vicinity of the WIPP to 10,000 years" (Larson, 2000). The assumption that the flow field is changing slowly and in equilibrium with the heads in the model domain is implicit in the suite of steady-state Culebra groundwater flow simulations used to assess the long-term performance of the WIPP. Recent Culebra pressure transducer data, shown in Figure 4, however, show highly episodic behavior and more than 25 m of drawdown in response to a single pumping well in, or at least with vertical connection to, the Culebra near the WIPP facilities. Figure 3 shows the location of the privately-owned groundwater well that is likely being pumped to supply the demand of the local Permian Basin hydrocarbon industry for water.

Long-term performance assessment modeling for radioactive waste disposal, unable to achieve the strictest interpretation of model validation (i.e., a post-audit), should demonstrate confidence, reasonable assurance, and adequacy/appropriateness (U.S. Congress, 1982). The objective of this study is to improve confidence in hydrologic analyses intended to provide quantitative predictions for the geologic timescale by evaluating the impact of unanticipated anthropogenic forcings on the Culebra groundwater flow system. Specifically, this simulation-based effort quantifies the influence of pumping at the private well on conservative Culebra particle paths and travel times from the map-view center of the repository waste panels to the LWA boundary (see Figure 3). The steps employed to meet the study objective include: (i) estimating a pumping rate for the private well, (ii) simulating drawdown associated with the pumping for a select, nine-month period, and (iii) simulating and comparing advective particle paths and travel times for cases with and without pumping. This study provides important information about the impact of land use change in the vicinity of the WIPP, motivation for better characterizing the storage properties of the Culebra, and elicits questions regarding validation of hydrologic models used in performance assessment of long-lived radioactive waste repositories.

## 1.1 The Culebra Dolomite

The Culebra Dolomite is a member of the Permian Rustler Formation in the Delaware Basin, the second largest subdivision of the Permian Basin. In early Permian time, shallow seas within the Delaware Basin dried into an extensive salt pan/saline mudflat in which cycles of carbonate, sulfate, and halite rock were deposited (Holt & Powers, 1988). The Salado Formation, which hosts the facility horizon for the WIPP (see Figure 1), was deposited at this time. Following this evaporation-dominated period, the depositional setting of the Delaware Basin shifted into a deeper and less saline, lagoonal-type environment. The Rustler Formation (see Figure 1 and Figure 2) reflects this environmental shift and the subsequent desiccation characterized by the deposition of carbonate, sulfate, and halite rock (Holt & Powers, 1998; Powers et al., 2006). Toward the end of Permian time, siltstones and sandstones of mixed shallow marine and fluvial origin (Holt & Powers, 1990) were deposited above the Rustler Formation as the Dewey Lake Formation (see Figure 2).

The Culebra, idealized in Figure 2, is an eight-meter thick dolomite/dolomitic limestone with an approximately one-degree eastward dip and is notable as the most transmissive and laterally extensive formation in the saturated zone above the repository (Powers et al., 2003). Saturated hydraulic conductivity of the Culebra, in the vicinity of the WIPP, varies by ten orders of magnitude and generally increases from east to west, reflecting regional trends in fracturing, secondary mineral infill, and porosity (Powers et al., 2003). As shown in Figure 3, confined groundwater flow in the Culebra near the WIPP is predominantly north-to-south. Knowledge of the Culebra groundwater flow field is largely attributable to an extensive monitoring network (see Figure 3), including 40 wells instrumented with transducers that report fluid pressure at 15-minute intervals with ± 0.2 m accuracy.

## 2.0 METHODS
## 2.1 Groundwater Flow Simulation

Initial conditions, physical parameters, boundary conditions, and a sink term were needed to solve the transient hydrogeologic boundary-value problems (BVPs) designed for this study. The boundary conditions and physical parameters correspond to the 100 two-dimensional, steady-state, saturated, heterogeneous, and anisotropic BVPs that currently serve as the foundation for Culebra-based performance assessment (Hart et al., 2009). The 100 east-west saturated hydraulic conductivity ($K_x$), north-south saturated hydraulic conductivity ($K_y$), and storativity ($S_s$) fields were generated by Hart et al. (2008, 2009) with a three-part approach. First, a conceptual model for geologic controls on hydrogeologic characteristics of the Culebra was formalized and regressed against measurements at monitoring wells. Second, the regression model was combined with a suite geologic maps to create 1,000 stochastic base realizations using the Geostatistical Software Library (Deutsch and Journel, 1997) and sequential indication simulation. Third, a pilot point calibration method was implemented using the inverse parameter estimation software PEST (Doherty, 2016) and the groundwater flow model MODFLOW2000 (Harbaugh, 2000). The 1,000 realizations were reduced to 100 realizations by Hart et al. (2009) based on model calibration against groundwater-levels in 2007 and nine planned multi-well pumping tests that pre-date the recent unplanned pumping activities. Although components of the BVPs used for Culebra-based performance assessment are employed here, the adequacy of the original BVPs is not evaluated as part of this research.

For this study, the 100 $K_x$, $K_y$, and $S_s$ realizations were geometrically averaged to form an ensemble average, resulting in 101 spatially variable fields with which to estimate a pumping rate for the private well. Figure 5 shows the distribution of spatially variable physical parameters for the average model. Summary statistics for the average model are provided in Table 1. Figure 3 shows the shape of the approximately 750 km$^2$ simulation domain and identifies the types/locations of boundary conditions. Identical constant head and no-flow boundary conditions, consistent with the Hart et al. (2009) approach, were applied across all of the realizations. The no-flow boundary follows a hydrologic divide along the axis of Nash Draw, a southwesterly trending topographic depression approximately 25 km long and five to 15 km wide. The eastern constant head boundary was specified as the surface elevation based on observations that the potentiometric surface in this area appears to roughly correspond with the topography (Hart et al., 2009). The northern, southern, and western constant head boundaries were assigned with a parametric surface equation (see Hart et al., 2009) that uses observed heads to interpolate values at the boundaries.

PFLOTRAN (Lichtner et al., 2015), a finite-volume groundwater flow and reactive solute transport code, was used to simulate two-dimensional, transient, saturated, heterogeneous, and anisotropic subsurface flow. The active region of the simulation domain (see Figure 3) consists of 57,240 square elements, 100 m on a side. Initial conditions were derived from steady-state simulation with no sink term. The simulation period spans approximately nine months from September 11$^{th}$, 2013 to June 1$^{st}$, 2014. The simulations reported here focus on this time interval because it encompasses the first and largest drawdown period visible in the Culebra monitoring network (see Figure 4). The rate at which the private well south of the LWA boundary (location shown in Figure 3) was pumped during this time is unknown.

For this study, a constant Culebra-based pumping rate subject to on-and-off periods was assumed. The pumping schedule was interpreted from a nearby observation well (i.e., H-4bR; see Figure 3) that is highly sensitive to drawdown in the private well. Sixteen drawdown observations, corresponding to the end of the nine-month period and the highest observed drawdown, were used as targets with which to maximize the Nash-Sutcliffe Model Efficiency ($E_f$) coefficient (Nash & Sutcliffe, 1970):

$$E_f = \frac{\sum_{i=1}^{n}(O_i - \bar{O})^2 - \sum_{i=1}^{n}(P_i - O_i)^2}{\sum_{i=1}^{n}(O_i - \bar{O})^2} \qquad (1)$$

where $P_i$ are the predicted drawdown values [m], $O_i$ are the observed drawdown values [m], n is the number of drawdown observations, and $\bar{O}$ is the mean of the observed drawdown values [m]. The $E_f$ is similar to the R$^2$ goodness-of-fit and, unlike the mean squared error or root mean squared error, measures performance with respect to a reference (Pebesma et al., 2005). An $E_f$ value of one indicates a perfect match between the model and the observed data. An $E_f$ value of zero indicates that the model predictions are as accurate as the mean of the observed data. An $E_f$ value less than zero indicates that the mean of the observed data is a better predictor than the model (Nash & Sutcliffe, 1970).

The observations chosen for this study correspond to locations where the rate of drawdown was readily apparent (i.e., at least 0.2 m month$^{-1}$). Individual pumping rates for the average model and the 100 original realizations were selected and simulated in an iterative fashion until a maximum $E_f$ value was achieved. Figure 6 shows the measured versus modeled values for each calibration target for the average model. The range of best-fit pumping rates and $E_f$ values calculated for the 100 realizations is shown in Figure 7. Figure 8 shows the drawdown field associated with the best-fit pumping rate for the average model and the 100 realizations.

## 2.2 Particle Tracking

The particle tracker code, DTRKMF (Rudeen, 2003), and the fluid pressures from the final time step of each of the 101 transient BVPs were used to calculate advective (i.e., non-dispersive and non-reactive) Culebra particle paths and travel times from the map-view center of the WIPP waste panels to the LWA boundary. DTRKMF tracks particles on a cell-by-cell basis using a semi-analytical solution, assuming that groundwater velocities vary linearly between the cell faces. The flow field used to parameterize DTRKMF does not change during the particle track simulation. Therefore, the particle paths and travel times provide an indication of the impacts related to continuous pumping activities at the private well. Travel times are faster than what would be expected in reality because DTRKMF does not account for mechanical dispersion, molecular diffusion, and sorption. Consideration of spreading and mixing processes would require numerical simulation of reactive solute transport and is beyond the scope of this study. To enable a comparison against particle tracks influenced by the pumping, the 101 steady-state groundwater flow fields used as initial conditions for the transient BVPs were also used for particle tracking. In total, 202 particle tracks were calculated. Figure 9 and Figure 10 show the influence of pumping on particle paths and travel times. Summary statistics from the simulations are provided in Table 2.

## 3.0 RESULTS
### 3.1 Groundwater Flow Simulation

The maximum $E_f$ value calculated for this study using the ensemble-averaged model is 0.7 and corresponds to a pumping rate of $1.9 \times 10^{-3}$ m$^3$s$^{-1}$ (30 gpm). Figure 6 is a plot of the measured versus modeled drawdown values for each observation location. In most cases, the measured versus modeled drawdown values fall within a five-meter envelope about the one-to-one (i.e., the perfect fit) line. The maximum $E_f$ value calculated among the 100 original realizations is 0.9 and corresponds to a pumping rate of $5.0 \times 10^{-3}$ m$^3$s$^{-1}$ (79 gpm). As shown in Figure 7, the best-fit pumping rates for the 100 realizations range from $4.4 \times 10^{-4}$ m$^3$s$^{-1}$ (7 gpm) to $6.2 \times 10^{-3}$ m$^3$s$^{-1}$ (98 gpm). The $E_f$ values range from -1.0 to 0.9. The average best-fit pumping rate across the 100 realizations is $2.4 \times 10^{-3}$ m$^3$s$^{-1}$ (38 gpm) with an $E_f$ value of 0.3.

The spatial distribution of drawdown associated with the best-fit pumping rate for the average model is shown in Figure 8a. The drawdown field is a roughly north-south trending lobe that broadens toward the south. A northeast-southwest band of low saturated hydraulic conductivity inside the LWA boundary (see Figure 5) results in tightly-spaced drawdown contours within the footprint of the repository. The south-central portion of the LWA boundary is strongly impacted by the pumping. Greater $K_x$ and $K_y$ variability in the southern portion of the BVP creates widely- and irregularly-spaced drawdown contours. Figure 6 and Figure 8a show that drawdown observations more accurately calculated by the model tend to cluster in the central portion of the LWA boundary. Consistent with observations in the field, the impacts of the private well do not appear to extend north beyond the LWA boundary during the nine months of pumping. The spatial distribution of drawdown associated with the best-fit pumping rate among the 100 realizations is shown in Figure 8b. Qualitatively, the general shape of the drawdown field for the average model and the individual realization are similar. The individual realization, however, shows more irregularly-shaped contours, half the drawdown at the private well, approximately twice the drawdown south of the well, and a drawdown footprint that is at least 24 % greater than the averaged model.

### 3.2 Particle Tracking

The simulated advective particle paths for the average model and realizations that comprise the average model, for cases with and without pumping, are shown in Figure 9. Summary particle travel distance statistics are provided in Table 2. Particles are tracked from a point in the Culebra directly above the map-view center of the WIPP waste panels to the LWA boundary. The particle tracks for all of the simulations are presented because they represent the range of possibility based on the realizations that currently serve as the foundation for Culebra-based performance assessment. Due to the predominantly north-south flow direction, all of the simulated particle tracks lead from the origin to the southern border of the LWA boundary, a straight line distance of 2.8 km. For the cases with no pumping, 76 % of the particles follow a high saturated hydraulic conductivity pathway near the southeastern corner of the LWA boundary (see Figure 5 and Figure 9a). The minimum, maximum, and average particle travel distances for the undisturbed (i.e., no pumping) cases are 3.0, 4.3, and 3.5 km, respectively. For the cases with pumping, 84 % of the particles track in a southwesterly direction toward the private well (see Figure 9b). The minimum, maximum, and average particle travel distances for the disturbed cases are 3.0, 5.5, and 3.7 km, respectively. Although pumping does not dramatically affect average particle travel distance, the activity does shift the intersection of the particle track with the LWA boundary more than two kilometers to the west.

The simulated advective particle travel times for the average model and realizations that comprise the average model, for cases with and without pumping, are shown in Figure 10. Summary particle travel time statistics are provided in Table 2. As mentioned previously, the flow fields corresponding to the end of the nine-month pumping periods were used to calculate particle travel times. Therefore, the particle paths and travel times provide an indication of the impacts related to continuous pumping activities at the private well. Sustained reductions in particle travel time would require steepened hydraulic gradients be maintained for decades, centuries, or even millennia. For the cases without pumping, the minimum, maximum, and average particle travel times are 2.6, 55.0, and 10.4 ka, respectively. For the cases with pumping, the minimum, maximum, and average particle travel times are 1.1, 20.7, and 4.5 ka, respectively. Despite the relatively low saturated hydraulic conductivity values along the average path for particles influenced by pumping, the average travel time is half that compared to the cases without pumping because of steepened hydraulic gradients. The pronounced increase in the frequency of shorter travel times for the cases with pumping, compared to those without pumping, is displayed in Figure 10.

## 4.0 DISCUSSION

The transient simulations conducted for this study provide estimates of pumping rates for the private well and permit an evaluation of model performance for the 100 spatially variable realizations of saturated hydraulic conductivity and storativity when subjected to an unplanned hydrologic forcing. Figure 7 shows a positive trend between $E_f$ values of -1.0 to 0.5 and best-fit pumping rates of 0 to $3.0 \times 10^{-3}$ m$^3$s$^{-1}$. This trend may reflect the limits of some realizations to transmit and store groundwater at rates than can accommodate the observed drawdown. Based on $E_f$ values greater than 0.5, the pumping rate is likely between $1.4 \times 10^{-3}$ m$^3$s$^{-1}$ (22 gpm) and $6.2 \times 10^{-3}$ m$^3$s$^{-1}$ (98 gpm). Detailed pumping records for the private well are not available, but an average annual rate of $4.1 \times 10^{-3}$ m$^3$s$^{-1}$ (66 gpm) (M. Schuhen, personal communication, December, 2$^{nd}$ 2016) suggests the estimates made by this study bound a realistic value.

Pumping activities at the privately-owned well in the vicinity of the WIPP constitute the most spatially and temporally extensive anthropogenic disturbance to groundwater levels recorded by the Culebra monitoring network to date. At least 43 % of the network shows evidence of drawdown. Based on the $1.9 \times 10^{-3}$ m$^3$s$^{-1}$ (30 gpm) average model pumping rate and the pumping schedule estimated by this study, approximately $3.3 \times 10^4$ m$^3$ (27 acre-ft) of water was extracted over nine months. For comparison, a typical Culebra hydraulic pumping test is limited to $3.7 \times 10^3$ m$^3$ (3 acre-ft) by the New Mexico Office of the State Engineer (DOE, 2009). The $1.9 \times 10^{-3}$ m$^3$s$^{-1}$ pumping rate corresponds to an $E_f$ value of 0.7. The intermediate degree of agreement between measured and modeled drawdown is unsurprising given that the spatially variable parameters (i.e., $K_x$, $K_y$, and $S_s$) were calibrated against stresses less than that imposed by the private well. Furthermore, $S_s$ is more poorly constrained than K$_x$ and K$_y$. As shown in Figure 5c, large swaths of the simulation domain are assigned a constant $S_s$ value because they are not situated in areas between former hydraulic pumping tests. The years of drawdown captured by the Culebra monitoring network, therefore, constitute a unique dataset amenable to inverse methods (e.g., Doherty, 2016) which could be used to better constrain spatial patterns of Culebra storage characteristics.

The work reported here underscores the potential in using knowledge gained over time to better inform WIPP-related analyses and the development of future repositories for radioactive waste. The steady-state simulations of groundwater flow currently used in Culebra-based performance assessment were calibrated against observations that pre-date pumping activities at the private well. Those steady-state groundwater flow fields are used to drive long-term transient radionuclide transport simulations as part of the compliance recertification for the WIPP (DOE, 2014). If the WIPP were being sited today, a set of observations markedly different than those associated with the pre-pumping activities would be used to formulate the steady-state groundwater flow fields. The particle paths and travel times for cases with pumping, shown in Figure 9b and Figure 10, provide a glimpse of how those alternative groundwater flow fields could impact predictions of radionuclide transport. A striking result of this study is the estimated doubling of conservative Culebra particle travel rates from a point above the map-view center of the waste panels to the LWA boundary. Pumping activity also shifts the intersection of the average particle path and the LWA boundary by more than two kilometers.

The observations and simulations presented by this study confirm that the Culebra groundwater flow system is not currently in steady state. In the future, transient groundwater flow simulation could be used to consider not only the intensity of future land use, but also its duration. For example, if the sink term used in this study is turned off at the end of the nine-month pumping period and the simulation allowed to proceed, particle paths and travel times return to a configuration similar to Figure 9a and Figure 10 within three years. The relatively short recovery timescale (years) compared to the average particle travel times (thousands of years) suggests that the effects of the nine-month pumping period are insignificant relative to the 10,000-year WIPP performance period.

Pumping activities at the private well, however, are not limited to a nine-month period. As shown in the H-4bR pressure transducer record (see Figure 4a), major pumping activities began in September 2013 and proceed in an episodic fashion through January 2016. Pumping at the private well continues at the time of writing. Transient anthropogenic forcings have impacted local Culebra fluid pressures for at least 18 % of the WIPP operational period thus far. Between 2010 and 2015, there has been an approximately six-fold increase in the annual development of oil and gas wells in the vicinity of the WIPP (Wagner & Thomas, 2016). It is possible that the demand for local sources of groundwater will persist and that other private wells will be developed (U.S. EIA, 2016). In such an environment, the Culebra may be expected to exhibit transient behavior for a growing proportion of the WIPP operational period. The location and intensity of land use associated with potential future wells is unknown and the pumping record focused on for this effort should not be interpreted as representative of future conditions.

## 5.0    SUMMARY AND CONCLUSIONS

Groundwater flow in the Culebra Dolomite Member (Culebra) of the Permian Rustler Formation is a potential pathway for the inadvertent release of radionuclides from the Waste Isolation Pilot Plant (WIPP). Subsurface flow models currently used to estimate the transport of radionuclides through the Culebra rely on groundwater flow fields calibrated against historical pumping tests and groundwater level observations made in 2007. In 2013, more than 40 % of the Culebra groundwater monitoring network began reporting highly episodic pressure fluctuations. The unprecedented magnitude of the drawdown, caused by pumping of groundwater from a nearby privately-owned well, may be driven by the demand of the local hydrocarbon industry for water.

In this study, numerical simulations of groundwater flow were used to estimate a pumping rate for the private well, simulate drawdown over a nine-month period, and simulate and compare advective particle paths and travel times for cases with and without pumping. The best-fit pumping rate calculated for an ensemble-averaged model is $1.9 \times 10^{-3}$ m$^3$s$^{-1}$ (30 gpm) and corresponds to a Model Efficiency ($E_f$) coefficient of 0.7. Best-fit pumping rates for the realizations that comprise the average model range from $4.4 \times 10^{-4}$ m$^3$s$^{-1}$ (7 gpm) to $6.2 \times 10^{-3}$ m$^3$s$^{-1}$ (98 gpm). Steepened hydraulic gradients generated by pumping activities at the private well shift the intersection of simulated particle paths with the point-of-compliance more than two kilometers to the west. The average particle travel time through the area-of-interest is halved from 10.4 to 4.5 ka, doubling the average rate of travel from 0.5 to 1.1 m yr$^{-1}$.

The hydrogeologic boundary-value problems offered here are the first to quantitatively illustrate that the Culebra groundwater flow system in the vicinity of the WIPP can be impacted by unplanned anthropogenic forcings. The nine-month pumping period that was the focus of this study does not result in conditions that pose a threat to long-term repository performance, but the intermediate level of agreement between measured and modeled drawdown suggests that improved parameterization of Culebra storage characteristics may be beneficial to future model development. The utility of such enhancements is highlighted by positive trends in the duration and intensity of land use in the vicinity of the WIPP.

This study demonstrates the ability of transient human-timescale simulation to improve confidence in steady-state, geologic-timescale simulation intended to represent long-term hydrologic response. In this case, a change in land use associated with groundwater resource development is observed to impose new stresses on the hydrologic system of interest. The perturbations are not thought to affect repository performance, but were, nevertheless, unexpected. It is unrealistic to assert that the type, intensity, and duration of every future transient perturbation can be anticipated over the timescales relevant to long-lived radioactive waste repositories. However, as demonstrated by this study, physics-based models driven with high-frequency (i.e., 15 minutes or less) groundwater monitoring data are capable of quantitatively addressing questions as they become important to those in the decision-management arena. High-frequency monitoring networks and flexible, physics-based simulation protocols will be increasingly appropriate for cases where the impacts of anthropogenic and natural phenomena must be deconvolved.

## ACKNOWLEDGEMENTS

This study would not have been possible without the herculean efforts of Michael D. Schuhen and the hydrology field team at Sandia National Laboratories in Carlsbad, New Mexico. The authors appreciate the thoughtful comments provided by Dr. Thomas F. Corbet and an anonymous reviewer on the submitted version of this manuscript. The first author is also grateful for the remarks provided by Dr. Ayman H. Alzraiee (U.S. Geological Survey), Professor Keith Loague (Leland Stanford Junior University), and Professor Theresa M. Schwartz (Allegheny College) on the first draft of this manuscript. Sandia National Laboratories is a multi-mission laboratory managed and operated by Sandia Corporation, a wholly owned subsidiary of Lockheed Martin Corporation, for the U.S. Department of Energy's National Nuclear Security Administration under contract DE-AC04-94AL85000. This research is funded by WIPP programs administered by the Office of Environmental Management (EM) of the U.S. Department of Energy.

**REFERNCES**

**Table 1.** Summary statistics for the spatially variable physical parameters used in the ensemble-averaged model.

| Parameter | Minimum | Maximum | Geometric mean | Variance |
|---|---|---|---|---|
| $\log(K_x)$ ms$^{-1}$ | -8.5 | -3.2 | -5.9 | -8.5 |
| $\log(K_y)$ ms$^{-1}$ | -8.7 | -2.8 | -5.9 | -7.8 |
| $\log(S_s)$ m$^{-1}$ | -6.3 | -2.3 | -5.1 | -5.6 |

$K_x$: east-west saturated hydraulic conductivity
$K_y$: north-south saturated hydraulic conductivity
$S_s$: specific storage

**Table 2**. Summary statistics for simulated advective particle paths and travel times from the Waste Isolation Pilot Plant to the Land Withdrawal Act boundary.

| Scenario | Travel distance [m] | | | Travel time [yr] | | |
|---|---|---|---|---|---|---|
| | Minimum | Maximum | Average | Minimum | Maximum | Average |
| No pumping | 2,967 | 4,291 | 3,465 | 2,626 | 54,977 | 10,437 |
| Pumping | 2,985 | 5,544 | 3,652 | 1,076 | 20,672 | 4,450 |

**Figure 1**

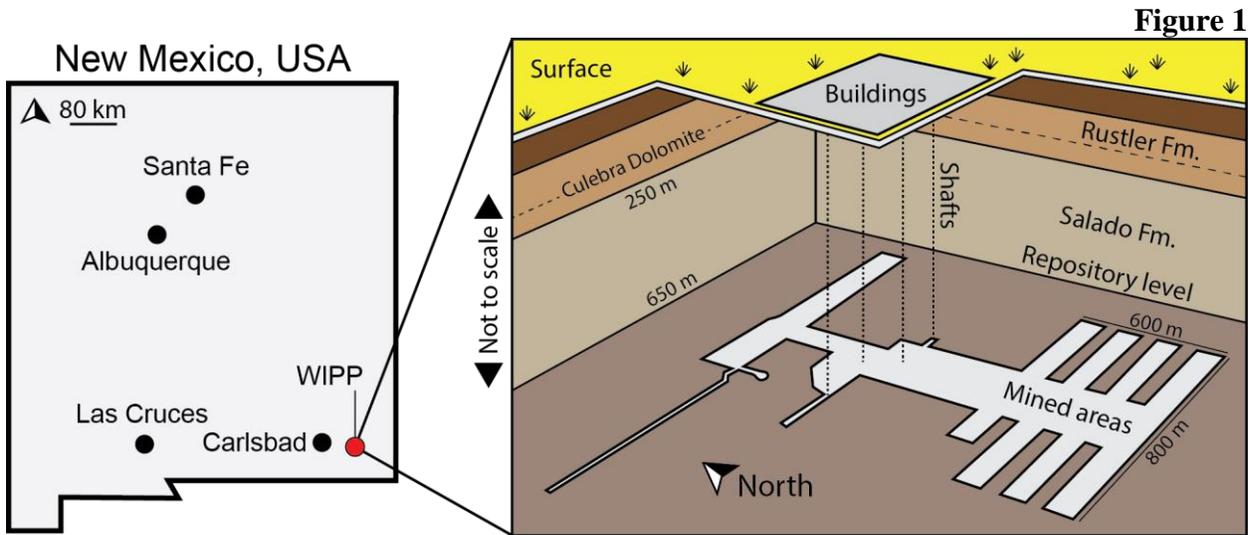

**Figure 1**. Location of the Waste Isolation Pilot Plant (WIPP) in southeastern New Mexico, USA and a cutaway schematic of the facility, including the surface infrastructure, shafts, and outline of mined areas.

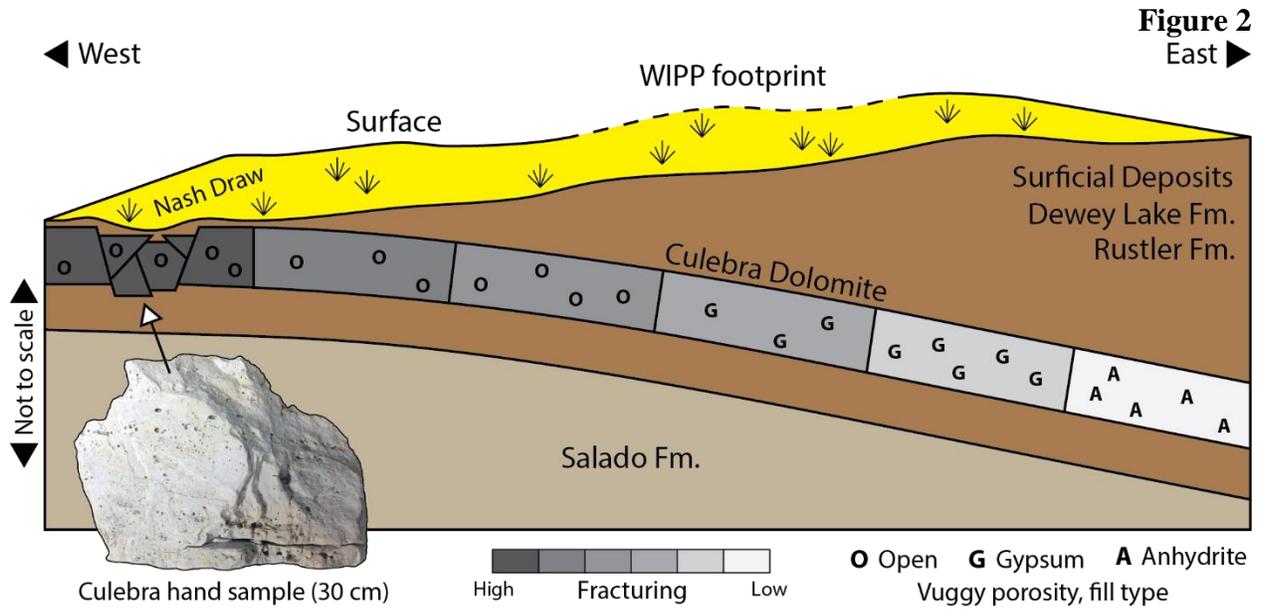

**Figure 2**. Conceptual cross section of the Culebra Dolomite Member (Culebra) of the Permian Rustler Formation in the vicinity of the WIPP. From east to west, vuggy porosity fill transitions from anhydrite, to gypsum, to open. Abundance and interconnectedness of fractures increases from east to west. Vertical scale and degree of eastward dip exaggerated for illustrative purposes. Culebra hand sample collected from the surface near Nash Draw. Adapted from Beauheim & Holt (1990).

**Figure 3**

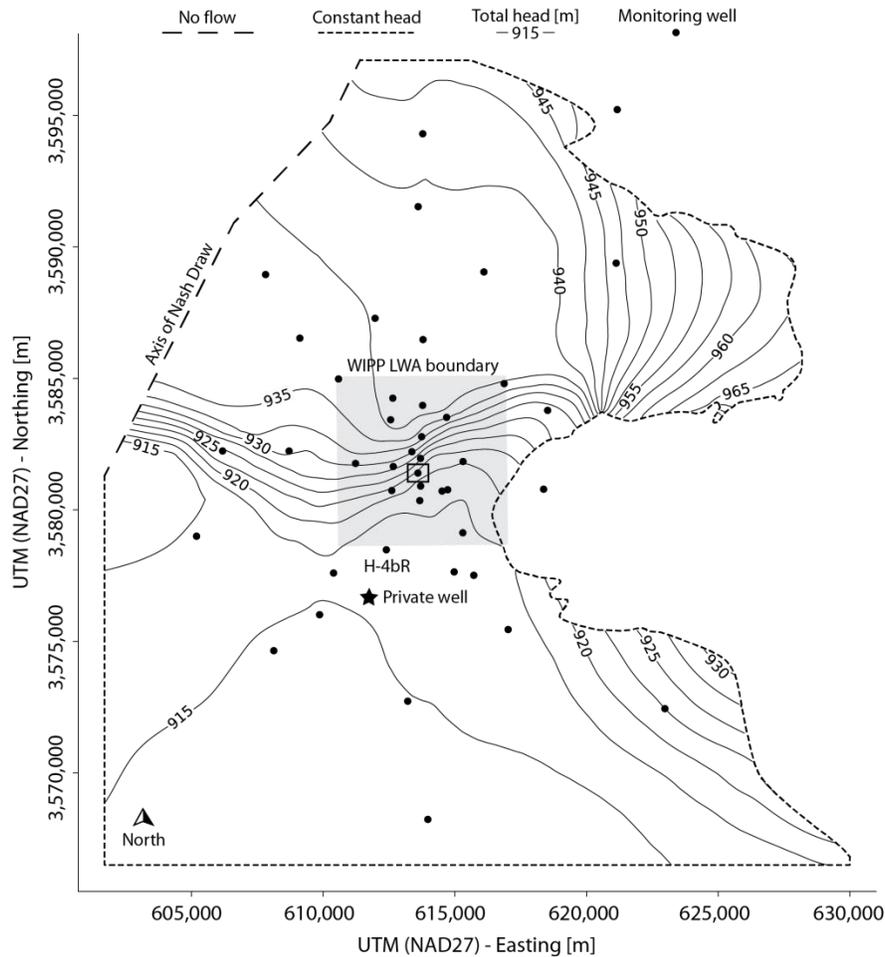

**Figure 3**. Culebra simulation domain with overlay of monitoring network and potentiometric contours calculated by Kuhlman (2014). Contours pre-date pumping activities at the private well. The gray square encompasses the 41.4 km$^2$ within the WIPP Land Withdrawal Act (LWA) boundary. The rectangle with a black border inside the LWA boundary encompasses the approximately 0.5 km$^2$ footprint of the mined waste panels. No-flow boundary corresponds to a topographic divide along the axis of Nash Draw. The eastern constant head boundary is specified as the surface elevation (Hart et al., 2009). The northern, southern, and western constant head boundaries are extrapolated from a parametric surface equation reported in Hart et al. (2009).

**Figure 4**

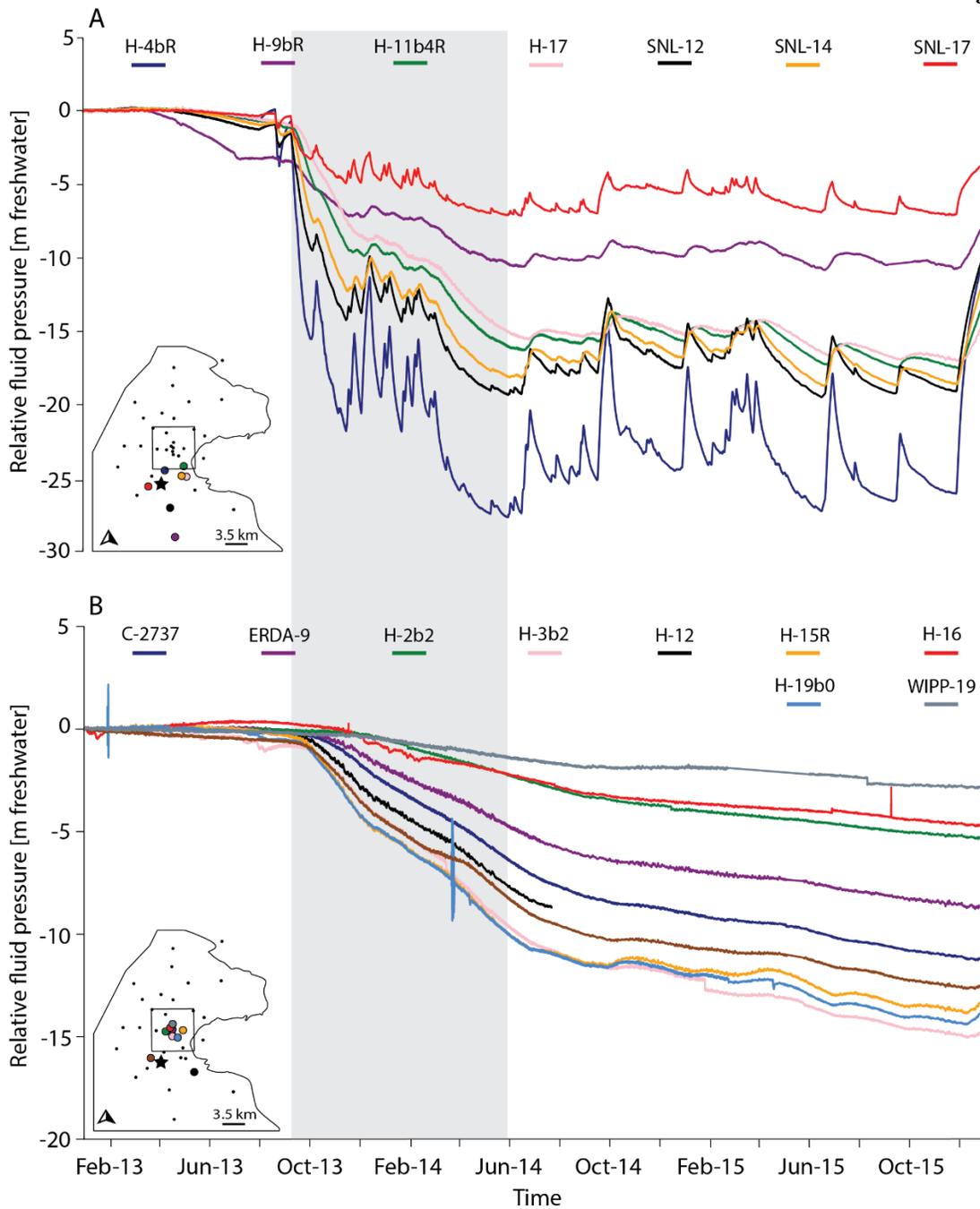

**Figure 4**. Pressure transducer data for Culebra observation wells exhibiting (a) strong and (b) subdued responses to pumping at the private well (location shown with black star). The nine-month period this simulation-based study focuses on is shaded in gray. The two pressure pulses visible in the H-19b0 record are due to pneumatic sinusoidal pulse tests.

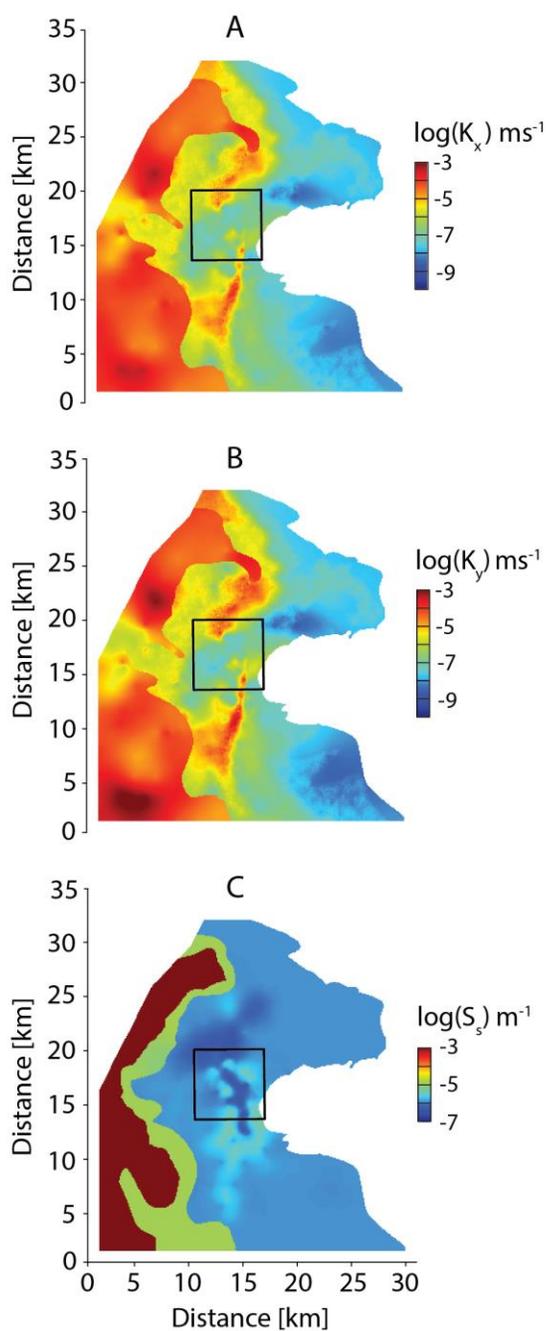

**Figure 5**. Map-view plots of the spatially variable distributions of (a) east-west saturated hydraulic conductivity ($K_x$), (b) north-south saturated hydraulic conductivity ($K_y$), and (c) specific storage ($S_s$) for the ensemble-averaged boundary value problem (BVP). Black squares correspond to the LWA boundary.

**Figure 6**

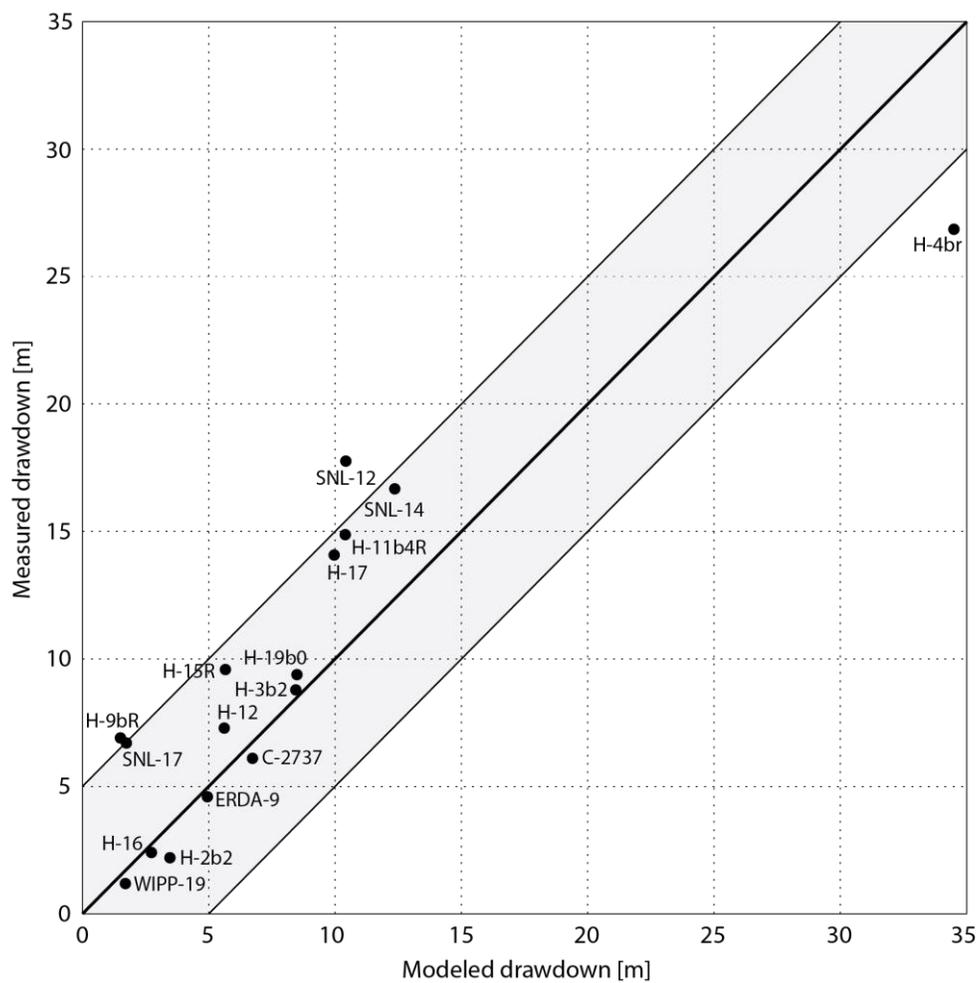

**Figure 6**. Best fit of measured versus modeled drawdown for the 16 observations used to calibrate the sink term for the ensemble-averaged BVP.

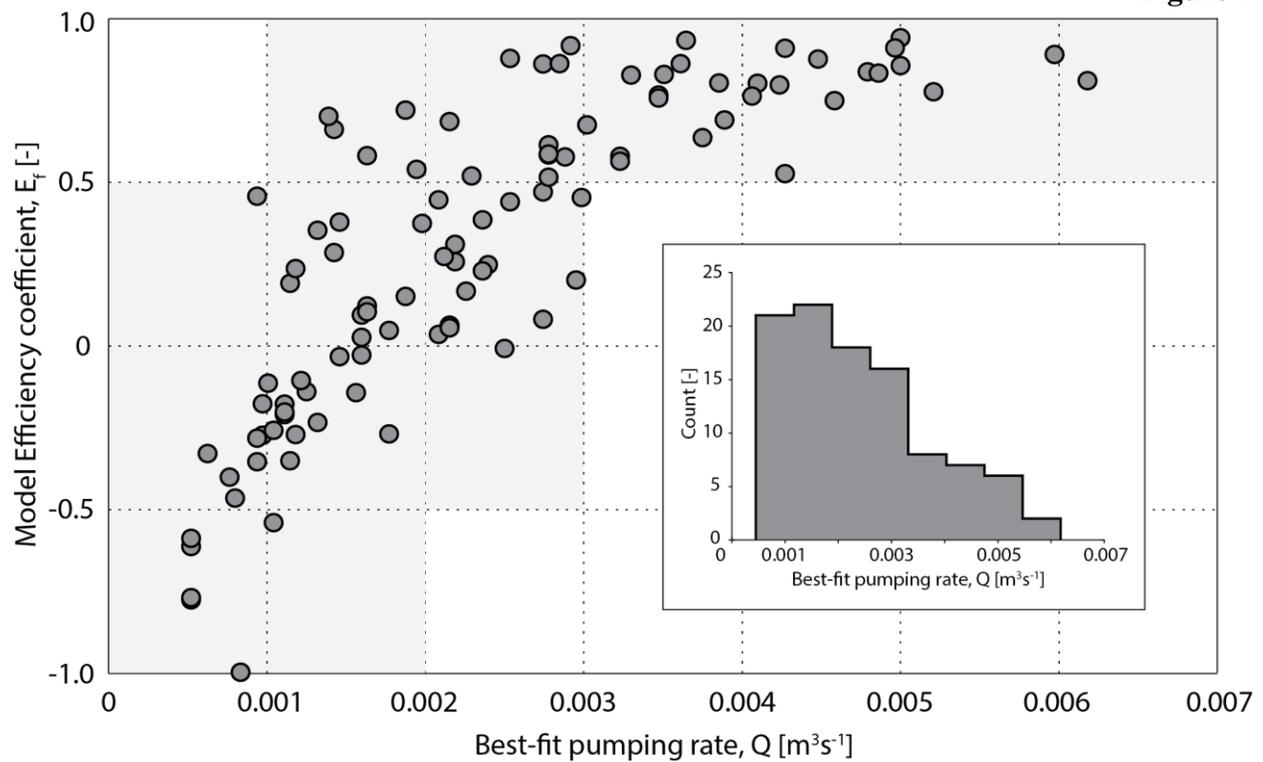

**Figure 7.** Scatter plot and histogram (inset) of the best-fit pumping rates (Q) for the 100 original realizations of $K_x$, $K_y$, and $S_s$. The average Q is $2.4 \times 10^{-3}$ m$^3$s$^{-1}$ (38 gpm) and the average Model Efficiency coefficient ($E_f$) is 0.3.

**Figure 8**

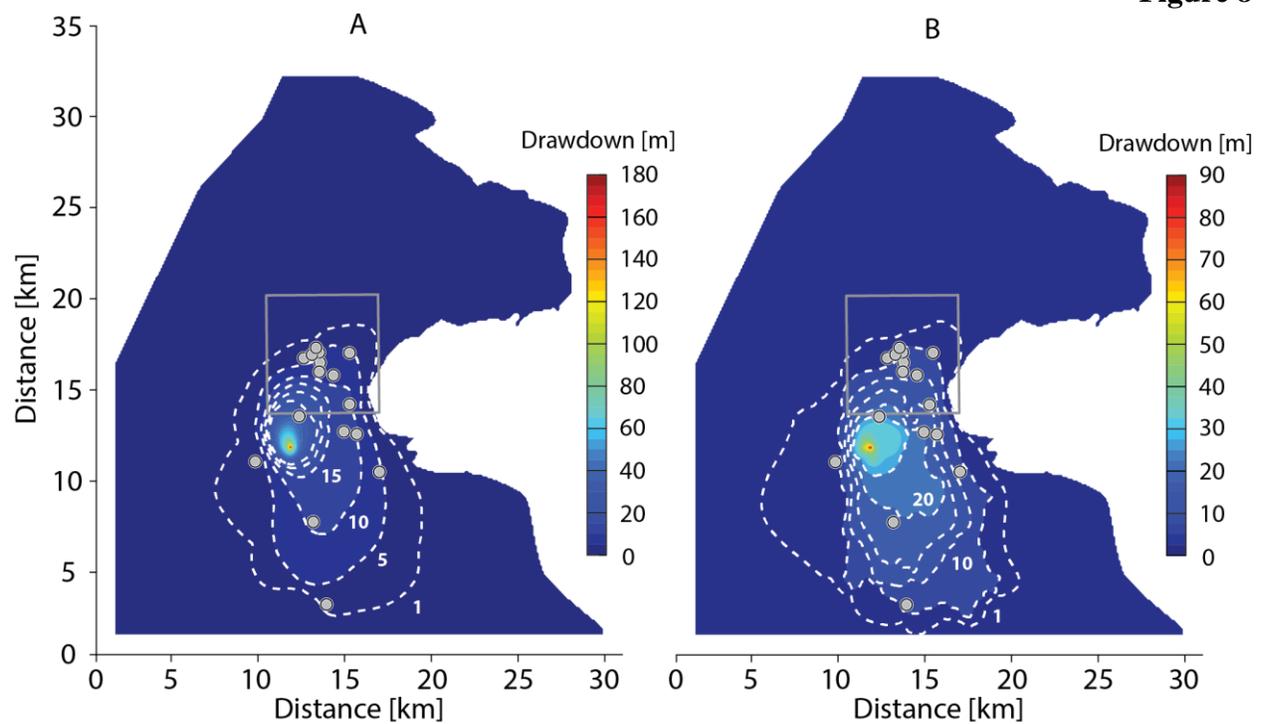

**Figure 8**. Plot of drawdown at the end of nine-month simulation period associated with the best-fit pumping rate (a) for the ensemble-averaged model and (b) among the 100 original realizations. White dashed lines used to distinguish the drawdown contours (i.e., 1 to 25 m) that approximately encompass the 16 observation locations. Gray square corresponds to the WIPP LWA boundary.

**Figure 9**

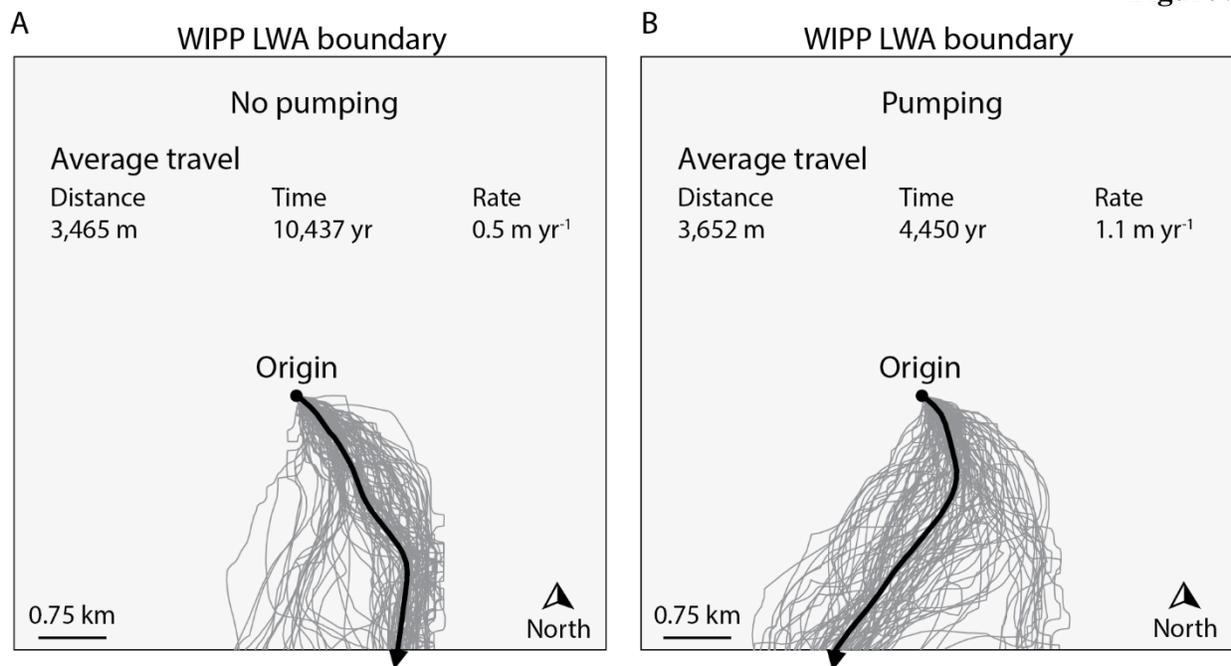

**Figure 9**. Simulated particle paths within the WIPP LWA boundary for the 100 original realizations of $K_x$, $K_y$, and $S_s$ (a) without and (b) with pumping at the private well. Thick black line corresponds to the particle path for the ensemble-averaged BVP.

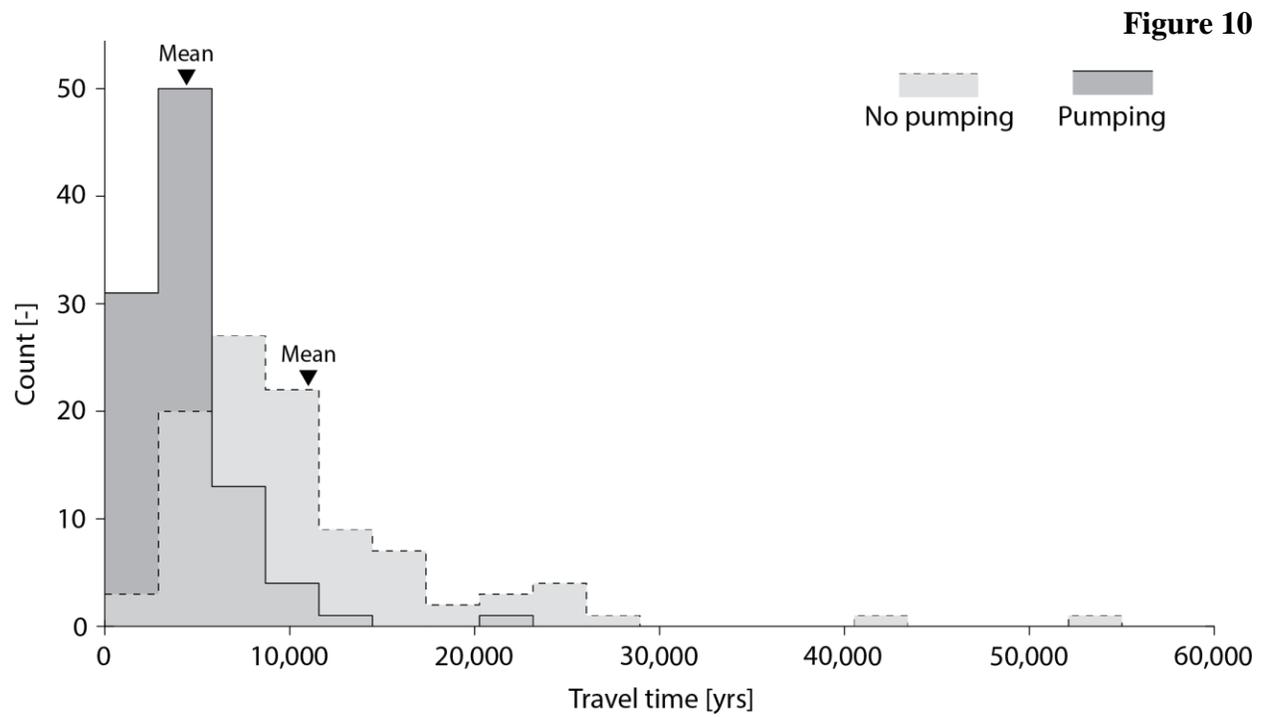

**Figure 10**. Histogram of simulated particle travel times for the 100 original realizations of $K_x$, $K_y$, and $S_s$ without and with pumping at the private well. The average travel times for the cases without and with pumping are 10.4 and 4.5 ka, respectively.